\documentclass[english,aps,prX,showpacs,twocolumn]{revtex4-2}
\usepackage[T1]{fontenc}
\usepackage[latin9]{inputenc}
\usepackage{babel}
\usepackage{graphicx}
\usepackage{bm}
\usepackage{bbm}
\usepackage{array}
\usepackage{amsmath}
\usepackage{amssymb}
\usepackage{dcolumn}
\usepackage{epstopdf}
\usepackage{bbold}
\usepackage{tikz}
\usepackage{pgfplots}
\usepackage[margin=15mm]{geometry}
\colorlet{linkequation}{blue}
\usepackage[colorlinks=true,linkcolor=blue,citecolor=blue,urlcolor=blue]{hyperref}
\usetikzlibrary{shadings}
\usepackage[nohyperlinks, nolist]{acronym}
\usepackage{braket}
\usepackage{mathrsfs}
\usepackage{import}
\usepackage{filecontents}
\usepackage{color}

\DeclareRobustCommand{\VAN}[3]{#3}

\begin{document}

\title{Skyrmions as quasiparticles: free energy and entropy
}

\author{Daniel Schick}
\affiliation{Fachbereich Physik, Universit\"at Konstanz, DE-78457 Konstanz, Germany}

\author{Markus Wei{\ss}enhofer}
\affiliation{Fachbereich Physik, Universit\"at Konstanz, DE-78457 Konstanz, Germany}

\author{Levente R{\'o}zsa}
\email[]{levente.rozsa@uni-konstanz.de}
\affiliation{Fachbereich Physik, Universit\"at Konstanz, DE-78457 Konstanz, Germany}

\author{Ulrich Nowak}
\affiliation{Fachbereich Physik, Universit\"at Konstanz, DE-78457 Konstanz, Germany}

\date{\today}

\renewcommand{\figurename}{FIG.}

\begin{abstract}
The free energy and the entropy of magnetic skyrmions with respect to the collinear state are calculated for a  (Pt$_{0.95}$Ir$_{0.05}$)/Fe bilayer on Pd(111) via atomistic spin model simulations. The simulations are carried out starting from very low temperatures where the skyrmion number is conserved up to the range where skyrmions are constantly created and destroyed by thermal fluctuations, highlighting their quasiparticle nature. The higher entropy of the skyrmions at low temperature leads to a reduced free energy, such that the skyrmions become energetically preferred over the collinear state due to entropic stabilization as predicted by linear spin-wave theory. Going beyond the linear spin-wave approximation, a sign change is shown to occur in the free energy as well as the entropy at elevated temperature. 
\end{abstract}



\maketitle

\begin{acronym}
\acro{DMI}[DMI]{Dzyaloshinsky-Moriya interaction}
\acro{sLLG}[sLLG]{stochastic Landau-Lifshitz-Gilbert}
\acro{LLG}[LLG]{Landau-Lifshitz-Gilbert}
\end{acronym}

\section{Introduction}
A magnetic spin configuration where the spin directions span the entire unit sphere is called a magnetic skyrmion \cite{Nagaosa2013,Mueller2015}.
Magnetic skyrmions were theoretically predicted to exist as solitons in the continuum two-dimensional isotropic Heisenberg model~\cite{Belavin1975}, but they were demonstrated to be destabilized by the external field, magnetocrystalline anisotropy or lattice discretization effects. Robust mechanisms for the stabilization of skyrmions were identified later, including the \ac{DMI} \cite{dmiD1958, dmiM1960, Bogdanov1994}, four-spin interactions~\cite{Heinze2011} and the frustration of Heisenberg exchange interactions~\cite{Okubo2012,vonMalottki2017}. The first experimental indications for the formation of a skyrmion lattice were found in MnSi via neutron scattering~\cite{Muelbauer2009}. Since then, skyrmions have been directly observed experimentally in other magnetic materials including Fe$_{1-x}$Co$_x$Si \cite{Yu2010,Muenzer2010}, Cu$_2$OSeO$_3$~\cite{Seki2012}, Pd/Fe/Ir(111)~\cite{Romming2013},  
and GaV$_4$S$_8$ \cite{Kezsmarki2015}. Skyrmions were even observed close to or at room temperature in Pt/Co/MgO nanostructures~\cite{Boulle2016}, thin films of FeGe~\cite{Yu2011} and Pt/Co/Ta~\cite{Woo2016}, and in Co-Zn-Mn alloys~\cite{Tokunaga2015}. 
Skyrmions are often regarded as exceptionally stable, turning them into a candidate for future use in logic and memory devices~\cite{Fert2013,Iwasaki2013,Zhou2014}. The operation of such devices relies on the demonstrated possibility of writing and deleting individual skyrmions~\cite{Hsu2017,Iwasaki2013}, and of moving them with electrical currents~\cite{Yu2012}.

Skyrmions are characterized by an integer topological charge $Q$, counting the number of times the spin configuration wraps the unit sphere. Because the topological charge cannot be changed dynamically in a continuum model, skyrmions are often referred to as topologically protected. 
However, the topological charge is not a conserved quantity in lattice spin models as the energy barrier between topologically trivial states and skyrmions is finite, allowing for the possibility of spontaneous creation and annihilation of skyrmions at finite temperature. Therefore, skyrmions should rather be thought of as quasiparticles, with their lifetime following the Arrhenius law as demonstrated in numerical simulations~\cite{Hagemeister2015,Rozsa2016} and experiments~\cite{Wild2017}. The decisive factor for skyrmion stability can be understood based on linear spin-wave theory or the harmonic approximation of the energy functional close to the metastable solution. In contrast to the collinear state, skyrmions are characterized by a number of low-frequency magnon modes which are easily excited by temperature, giving rise to a higher spin-wave entropy and, consequently, a free-energy preference for skyrmions over collinear configurations as the temperature is raised ~\cite{Muelbauer2009}. The larger entropy means that the attempt frequency, i.e. the pre-exponential factor in the Arrhenius law, is lower for skyrmion annihilation than for the creation of a skyrmion from the collinear state~\cite{Hagemeister2015, Desplat2018, Malottki2019}, referred to as entropic stabilization.
However, linear spin-wave theory is expected to gradually lose its validity at elevated temperatures, where magnon-magnon interactions become more prominent and the fast creation and destruction of skyrmions makes the expansion around a well-defined equilibrium state questionable. Numerical simulations are a suitable method for investigating magnetic systems beyond the linear spin-wave approximation, but extracting the free energy or the entropy is typically challenging since these quantities are not defined for single microstates. Therefore, the thermodynamic properties of skyrmions beyond the linear spin-wave approximation remain to be explored.

In this paper, we calculate the free-energy and entropy difference between topologically trivial and skyrmionic states in a wide temperature range through numerical simulations for a (Pt$_{0.95}$Ir$_{0.05}$)/Fe bilayer on a Pd(111) surface. The free-energy difference is shown to decrease with temperature, leading to a range where the free energy of skyrmionic states with $Q=1$ is lower than for topologically trivial states. The dependence of this temperature range on the magnetic field and the system size is demonstrated. Remarkably, we also find a temperature range where skyrmions possess a lower entropy than topologically trivial states, reversing the entropic stabilization. These results highlight the thermodynamic properties and the quasiparticle character of skyrmions.

\section{Methods} 
\subsection{Spin dynamics simulations}
The system being modeled is a (Pt$_{0.95}$Ir$_{0.05}$)/Fe bilayer on a Pd(111) surface. The magnetic Fe moments are described by the following classical atomistic Hamiltonian: 
\begin{align}
\mathcal{H} = \frac{1}{2} \sum\limits_{i \neq j} \bm{S}_i \mathcal{J}_{ij} \bm{S}_j + \sum\limits_{i} \bm{S}_i \mathcal{K} \bm{S}_i  - \mu_{\textrm{s}} \sum\limits_i \bm{B} \cdot \bm{S}_i.
\end{align}
Here, $i,j$ are site indices and $\mu_{\textrm{s}}$ is the spin magnetic moment. The interaction tensors $\mathcal{J}_{ij}$ include Heisenberg exchange in its diagonal terms $J_{ij}=\dfrac{1}{3}\textrm{Tr}\mathcal{J}_{ij}$, DMI in its antisymmetric part $\bm{D}_{ij}\left(\bm{S}_{i}\times\bm{S}_{j}\right)=\dfrac{1}{2}\bm{S}_{i}\left(\mathcal{J}_{ij}-\mathcal{J}^{T}_{ij}\right)\bm{S}_{j}$, and two-site anisotropy in its traceless symmetric part. 
$\mathcal{K}$ denotes the on-site anisotropy tensor. $\bm{B}$ is an applied external field perpendicular to the surface. The exchange coefficients $\mathcal{J}_{ij}$ and $\mathcal{K}$ were determined by {\it ab initio} calculations using the Korringa--Kohn--Rostoker~\cite{Szunyogh1995, Zeller1995} multiple scattering formalism with the relativistic torque~\cite{Udvardi2003} method, and are reported in Refs.~\cite{Rozsa2017,Zazvorka2019}. The interactions were included between pairs of spins up to a distance of $8$ lattice constants. In particular, the competition between ferromagnetic nearest-neighbor and antiferromagnetic next-nearest-neighbor interactions leads to the stabilization of localized spin structures with various topological charges, including skyrmions and antiskyrmions~\cite{Rozsa2017,Weissenhofer2019,Weissenhofer2020,Rozsa_2020}.

The dynamics of the spin system is described by the stochastic Landau--Lifshitz--Gilbert (LLG) equation,
\begin{align}
\frac{\partial \bm{S}_i}{\partial t} = - \frac{\gamma}{(1+\alpha^2) \mu_{\textrm{s}}} \bm{S}_i \times (\bm{H}_i + \alpha \bm{S}_i \times \bm{H}_i),
\end{align}
where $\alpha$ is the damping parameter and $\gamma$ the gyromagnetic ratio. $\bm{H}_i$ is the local effective field with $\bm{H}_i = \bm{\zeta}_i - \partial \mathcal{H} / \partial \bm{S}_i$. $\bm{\zeta}_i$ is a Gaussian noise term with $\langle \bm{\zeta}_i(t) \rangle = 0$ and $\langle \zeta_{i,\mu}(t) \zeta_{j,\nu} ( t^{\prime}) \rangle = \delta_{i,j} \delta_{\mu,\nu} \delta(t-t^{\prime}) 2 \alpha k_B T \mu_s / \gamma$, with $i,j$ denoting different spins and $\mu,\nu$ representing different Cartesian coordinate directions.

The simulations were performed on a two-dimensional triangular lattice with periodic boundary conditions. The damping constant was set to $\alpha=1$ in order to increase the speed of relaxation towards thermal equilibrium and to make transitions between states with different topological charges more frequent. 
The simulations were started from a collinear field-polarized state or from a configuration with a prepared skyrmionic structure of given topological charge and the time evolution was calculated according to the stochastic LLG equation for a time of $50.4 \, \mu\textrm{s}$ with a total of 5 independent realizations for each temperature and initial condition. The time step of the simulation was set to $50.4 \, \textrm{fs}$ and the topological charge was calculated at each step.

\subsection{Calculation of thermodynamic quantities}
The free energy and the entropy of a skyrmion are not defined for single microstates of the system; therefore, they cannot be calculated as time or ensemble averages. Instead, they were determined from the time dependence of the topological charge and the energy of the system at various values of the external parameters.

The topological charge of a continuous vector field can be calculated as a surface integral with the spin vectors $\bm{S}$ of unit length,
\begin{align}
    Q= -\frac{1}{4\pi} \int \mathrm{d}^2r \, \bm{S} \cdot (\partial_x \bm{S} \times \partial_y \bm{S}).\label{eqn3}
\end{align}
The sign convention is chosen such that a skyrmion on an out-of-plane-oriented collinear background has a topological charge of $Q=1$, while an antiskyrmion is described by $Q=-1$. For our spin model simulation we use a discretized version of Eq.~\eqref{eqn3}~\cite{Boettcher2018}, calculating $Q$ via 
\begin{align}
    Q\left(\bm{S}\right) = -\sum_{\left\{i,j,k\right\}}\frac{1}{2\pi} \arctan \left( \frac{\bm{S}_i \cdot \left(\bm{S}_j\times \bm{S}_k\right) }{1+ \bm{S}_i \cdot \bm{S}_j + \bm{S}_i \cdot \bm{S}_k + \bm{S}_j \cdot \bm{S}_k} \right),
\end{align}
where $\bm{S}_i$ ($i=1,2,3)$ denotes the spin unit vectors on a nearest-neighbor triangle on the lattice. This value is calculated for all simulated triangles and summed up to determine the topological charge for the entire system. For the simulated system with periodic boundary conditions, $Q$ is always an integer value.

To each value of the topological charge $Q$ we assign the conditional free energy $F_Q$. This can be connected to the conditional partition function $Z_Q$, defined as the phase-space integral of the Boltzmann exponential factor over all configurations with topological charge $Q$,
\begin{align}
    F_Q = - k_{\textrm{B}} T \ln(Z_Q) = - k_{\textrm{B}} T \ln \left( \int_{Q\left(\bm{S}\right)=Q} \exp\left(-\beta\mathcal{H}\left(\bm{S}\right)\right) \, \mathrm{d}\bm{S} \right),
\end{align}
with $\beta = 1/\left(k_\textrm{B} T\right)$. The direct calculation of the conditional partition functions is numerically not feasible. However, the difference in free energies between two values of the topological charge, for example $1$ and $0$, can be calculated from the ratio of partition functions using the corresponding condition:
\begin{align}
    \Delta F_{10} = F_1 - F_0 = - k_{\textrm{B}} T \ln( Z_1/Z_0).\label{eqn6}
\end{align}
During a numerical simulation, a total number $N$ of spin configurations, or recorded events, are created. Once the system has reached thermal equilibrium, the configurations are generated with the probabilities according to the Boltzmann distribution. Therefore, we expect the number of recorded events $N_Q$ fulfilling condition $Q$ divided by the total number of recorded events $N$ to converge to the ratio of the partition function $Z_Q$ and the total partition function $Z$ in the thermodynamic limit: 
\begin{align}
    \lim_{N\rightarrow\infty}N_Q / N = Z_Q / Z.\label{eqn7}
\end{align}
Using Eqs.~\eqref{eqn6} and \eqref{eqn7}, a formula to calculate free-energy differences can be derived, only based on the number of recorded events during simulation time that fulfill a certain condition \cite{vanGunsteren2002}, 
\begin{align}
\Delta F_{10,\textrm{count}}(T) = \lim_{N\rightarrow\infty} - k_{\textrm{B}} T \ln( N_1 / N_0) = - \Delta F_{01,\textrm{count}}(T), \label{DeltaFN}
\end{align}
which we apply to spin configurations with different topological charges $Q$. Equation~\eqref{DeltaFN} requires that the simulation explores a large part of the phase space, such that a sufficient amount of changes in the topological charge take place during simulation time, with both states being present at a considerable number of time steps.

The lifetime of configurations with a given value of the topological charge is expected to follow the Arrhenius law $\tau~\propto\textrm{exp}\left(\Delta E/\left(k_{\textrm{B}} T\right)\right)$, where $\tau$ is the lifetime of the state and $\Delta E$ is the energy barrier separating it from a configuration with a different topological charge. The lifetimes in the simulations have been measured using the method described in Ref.~\cite{Rozsa2016}, and the energy barriers were estimated from an Arrhenius fit to the data. These energy barriers were compared to values obtained from the Geodesic Nudged Elastic Band (GNEB) method~\cite{Bessarab2015} as implemented in the UppASD simulation code~\cite{UppASD}. The GNEB method provides an analytic approximation for the energy barrier and has been used for investigating skyrmion lifetimes in numerous works previously~\cite{Bessarab2018, Desplat2018, Malottki2019}.

At lower temperatures, switching events between configurations with different topological charges become exceedingly rare. In the regime where no changes in topological charge take place during the simulation, another method must be employed to calculate the free energy. Using the internal energy $U_{Q}=\left<\mathcal{H}\left(\bm{S}\right)\right>_{Q\left(\bm{S}\right)=Q}$ and a single value of $\Delta F_{10}$ at a high temperature $T_0$ calculated from Eq.~\eqref{DeltaFN}, $\Delta F_{10}$ can be determined at all temperatures based on the internal energy difference $\Delta U_{10}=U_{1}-U_{0}$ via \cite{vanGunsteren2002,Hinzke2008},
\begin{align}
\Delta F_{10,\textrm{integ}}(T) = \Delta F_{10,\textrm{count}}(T_0) \frac{T}{T_0} - T \int_{T_0}^T \Delta U_{10}(T^\prime) \frac{\mathrm{d}T^\prime}{T^{\prime 2}}. \label{DeltaFInteg}
\end{align}
$\Delta U_{10}$ is easily accessible as the time average of the Hamiltonian, which may be determined from independent simulations initialized in states with different topological charges, rather than requiring a high number of switching events during a single run. Note that Eq.~\eqref{DeltaFInteg} may still become numerically inaccurate close to $T=0$ where the denominator goes to zero.

The entropy difference between skyrmionic and topologically trivial states can be calculated by the negative derivative of the free energy difference,
\begin{align}
\Delta S_{10}=-\frac{\partial\Delta F_{10}}{\partial T}.\label{eqn9}
\end{align}
which is done numerically by using the central finite difference.

The average value of the topological charge can be calculated as a time average of $Q$ using the number of recorded events with corresponding topological charge:
\begin{align}
    \langle Q \rangle_{\textrm{time}} = \frac{\sum_Q Q N_Q }{\sum_Q N_Q}.\label{Qtime}
\end{align}
This method of calculation is appropriate at high temperatures. However, at low temperatures there are no changes in topological charge during the simulation time, making the calculated value dependent on the initial condition.

For the low-temperature calculations, we use the formula
\begin{align}
    \langle Q \rangle_{\textrm{ensemble}} = \frac{\sum_{Q}Q\exp(-\beta \Delta F_{Q0})}{\sum_{Q}\exp(-\beta \Delta F_{Q0})},\label{Qens}
\end{align}
where $\Delta F_{Q0}$ is determined from Eq.~\eqref{DeltaFInteg}. Note that here the free-energy difference of the entire system has to be used and not an average per spin. Since this method requires performing simulations with different initial values of $Q$, we refer to it as an ensemble average. For simplicity, the summation in Eq.~\eqref{Qens} were restricted to the values $Q=-1,0,1$, which is possible since skyrmions and antiskyrmions are both stable at zero temperature in the system~\cite{Rozsa2017}. Calculating the free-energy difference for other values of the topological charge is numerically demanding, and their relative Boltzmann weight is significantly lower. When the number of states with other topological charges increases, this restriction of the sum to $Q=-1,0,1$ is no longer a good approximation, and Eq.~\eqref{Qtime} can be used directly instead.

\subsection{Linear spin-wave theory\label{sec2c}}

At sufficiently low temperatures, the internal and the free energy, as well as the entropy may be determined analytically based on linear spin-wave expansion. In this approximation, the conditional free energy may be expressed as~\cite{Rozsa2015}
\begin{align}
F_{Q}=U_{Q}-TS_{Q},\label{eqn10}
\end{align}
with
\begin{align}
U_{Q}=E_{Q}+\sum_{k=1}^{N_{S}}\frac{\mu_{\textrm{s}}}{\gamma}\omega_{Q,k}n_{Q,k}\label{eqn11}
\end{align}
and
\begin{align}
S_{Q}=k_{\textrm{B}}\sum_{k=1}^{N_{S}}\textrm{ln}\:n_{Q,k},\label{eqn12}
\end{align}
with $E_{Q}$ the energy of the equilibrium spin configuration at zero temperature and $\omega_{Q,k}$ denoting the spin-wave frequencies in the vicinity of the local energy minimum. This method corresponds to approximating the energy landscape close to the minimum with the potential of independent harmonic oscillators. The calculation of the spin-wave frequencies for non-collinear spin configurations is discussed in, e.g., Ref.~\cite{Rozsa2018}.

The occupation number of the spin-wave modes according to classical statistics is calculated as
\begin{align}
n_{Q,k}=\frac{\int q^{2}_{Q,k}\textrm{e}^{-\beta \frac{\mu_{\textrm{s}}}{\gamma}\omega_{Q,k} q^{2}_{Q,k}}\textrm{d}q_{Q,k}}{\int \textrm{e}^{-\beta \frac{\mu_{\textrm{s}}}{\gamma}\omega_{Q,k} q^{2}_{Q,k}}\textrm{d}q_{Q,k}}=\frac{k_{\textrm{B}}T}{\frac{\mu_{\textrm{s}}}{\gamma}\omega_{Q,k}}\:\:\:\textrm{for}\:\:\:\omega_{Q,k}\neq 0,\label{eqn12a}
\end{align}
where the integration limits for the phase space variable $q_{Q,k}$ are extended to infinity. For eigenmodes with $\omega_{Q,k}=0$, the harmonic-oscillator approximation loses its validity. Such Goldstone modes naturally occur for skyrmions in the continuum limit, since translation along one of the two in-plane directions does not influence the energy of the system. As discussed in Ref.~\cite{Bessarab2018}, the phase-space variable belonging to the translational modes can be expressed as
\begin{align}
q_{Q,\mu}=&A_{Q,\mu}r_{\mu},\label{eqn12b}
\\
A_{Q,\mu}=&\left|\partial_{\mu}\boldsymbol{S}^{Q}\right|=\sqrt{\sum_{i=1}^{N_{S}}\left(\partial_{\mu}\boldsymbol{S}_{i}^{Q}\right)^{2}},\label{eqn12c}
\end{align}
where $\mu=x,y$ denotes translation along one of the in-plane coordinates $r_{\mu}$. Substituting Eqs.~\eqref{eqn12b} and \eqref{eqn12c} into Eq.~\eqref{eqn12a} yields
\begin{align}
n_{Q,\mu}=\frac{\int q^{2}_{Q,\mu}\textrm{d}q_{Q,\mu}}{\int \textrm{d}q_{Q,\mu}}=\frac{A_{Q,\mu}^{2}L_{\mu}^{2}}{3},\label{eqn12d}
\end{align}
with $L_{\mu}$ the system size along the given direction. Since the zero as well as the nonzero eigenvalues enter Eq.~\eqref{eqn11} in pairs, the occupation number for the pair of translational modes will be written as $n_{Q,k=1}=\sqrt{n_{Q,x}n_{Q,y}}$.

Equations~\eqref{eqn10}-\eqref{eqn12} imply $\Delta F_{10}=\Delta U_{10}$ at $T=0$~K. Since $\frac{\mu_{\textrm{s}}}{\gamma}\omega_{Q,k}n_{Q,k}=k_{\textrm{B}}T$ is independent of the spin configuration for finite-frequency modes, one has 
$\Delta U_{10}=\Delta U_{10}\left(T=0\right)-k_{\textrm{B}}T$ because of the translational mode of the skyrmion. For the entropy difference one obtains
\begin{align}
\frac{1}{k_{\textrm{B}}}\Delta S_{10}=\sum_{k=1}^{N_{S}}\textrm{ln}\:\omega_{0,k}-\sum_{k=2}^{N_{S}}\textrm{ln}\:\omega_{1,k}-\textrm{ln}\frac{k_{\textrm{B}}T}{\frac{\mu_{\textrm{s}}}{\gamma}n_{1,1}}.\label{eqn13}
\end{align}
The main contribution to the temperature dependence of $\Delta F_{10}$ comes from the entropy contribution of the finite-frequency modes, leading to a linear decrease. A logarithmic divergence of the entropy difference due to the Goldstone modes of the skyrmion is predicted from Eq.~\eqref{eqn13} at low temperatures, an effect also observed in the calculation of the configurational entropy of skyrmions~\cite{Zivieri2019}. In the lattice model considered here, this divergence is regularized since the atomic sites create a weak but finite periodic modulation potential for the skyrmion position, and the translational modes obtain a finite frequency. In Ref.~\cite{Desplat2018}, it was argued that the translational modes in a lattice model can be treated like all other spin-wave excitations, rather than Goldstone modes, during the calculation of skyrmion lifetimes. The logarithmic singularity should be less pronounced for larger system sizes, since the number of Goldstone modes does not scale with the number of spins. 

Linear spin-wave theory is expected to break down as the temperature becomes comparable to the energy barrier separating the different metastable equilibrium states. For the translation of skyrmions, this effect is observable already at the energy scales corresponding to the atomic modulation potential of the skyrmion position, above which a Brownian motion of the quasiparticles can be observed~\cite{Zazvorka2019,Weissenhofer2020}. Overcoming the energy barrier between the skyrmion and the topologically trivial state requires considerably higher temperatures. Finally, increasing the temperature also enhances the role of magnon--magnon interactions as the spin-wave occupation numbers become higher, further limiting the applicability of the linear approximation.

\section{Results} 

\begin{figure}
\includegraphics[width=0.9\columnwidth]{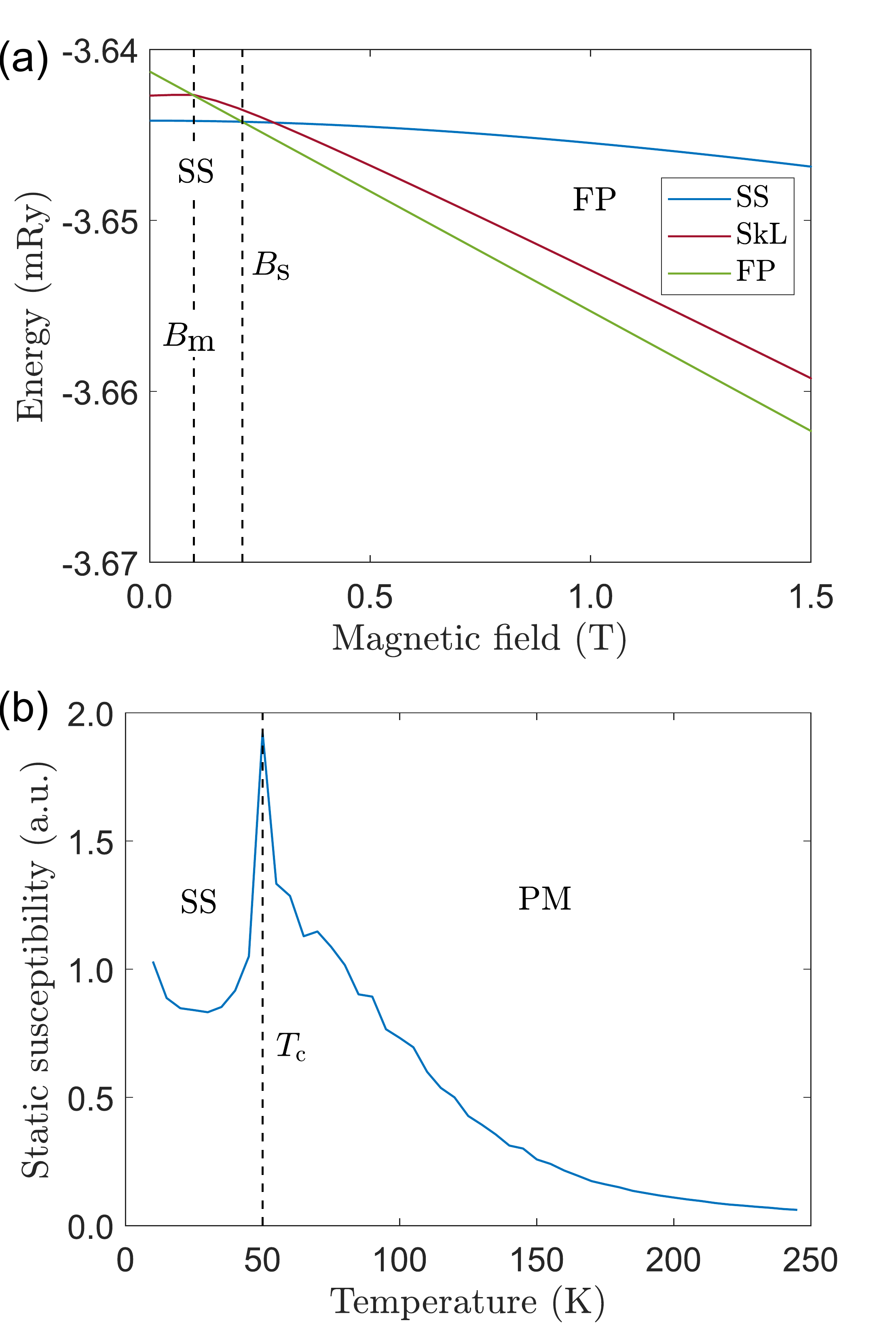}
\caption{Phase diagram of the (Pt$_{0.95}$Ir$_{0.05}$)/Fe/Pd(111) system. (a) Energy per spin in the spin spiral (SS), skyrmion lattice (SkL) and field-polarized (FP) states as a function of magnetic field at $T=0\,\textrm{K}$. $B_{\textrm{m}}$ and $B_{\textrm{s}}$ denote the metastability field for an isolated skyrmion and the transition field from the spin spiral to the field-polarized state, respectively. (b) Static susceptibility as a function of temperature at $B=0\,\textrm{T}$. $T_{\textrm{c}}$ denotes the transition temperature from the SS to the paramagnetic (PM) phase.
}
\label{phasediag}
\end{figure}

Before discussing the thermodynamic properties of metastable skyrmions, we present the observed phases in the system. The zero-temperature energies of the different states are displayed in Fig.~\ref{phasediag}(a). For lower external fields the ground state is a spin spiral which transforms into a collinear field-polarized state at around $B_{\textrm{s}}\approx0.21$~T~\cite{Rozsa2016v2}. The skyrmion lattice is not a ground state for any value of the external field, since its energy already exceeds that of the field-polarized state at $B_{\textrm{s}}$. A single isolated skyrmion leads to a positive energy contribution to the field-polarized state for fields above $B_{\textrm{m}}\approx0.1$~T. Due to the short-range attractive interaction between the skyrmions, this field is slightly lower than where the energy curves for the skyrmion lattice and the field-polarized states cross in Fig.~\ref{phasediag}(a). In the following, we study the thermal properties of metastable skyrmions at field values higher than $B_{\textrm{m}}$.

As the temperature is increased, the spin spiral phase is transformed into the paramagnetic phase at $T_{\textrm{c}}\approx50$~K for $B=0$~T, as evidenced by the singular behaviour of the static magnetic susceptibility shown in Fig.~\ref{phasediag}(b). Based on previous studies~\cite{Rozsa2016,Boettcher2018}, the critical temperature only changes weakly as the field is increased to $B_{\textrm{s}}$. The field-polarized regime is part of the paramagnetic phase, with a continuous crossover in the physical observables as the temperature is increased. Above the critical temperature, strong thermal fluctuations lead to the formation of a considerable number of metastable skyrmions in an intermediate regime reaching approximately up to $120$~K, where an inflection point in the static susceptibility is found. This intermediate regime extends to considerably higher field values than $B_{\textrm{s}}$, as has been investigated for similar systems in Refs.~\cite{Rozsa2016,Boettcher2018}.

\begin{figure}
\includegraphics[width=0.9\columnwidth]{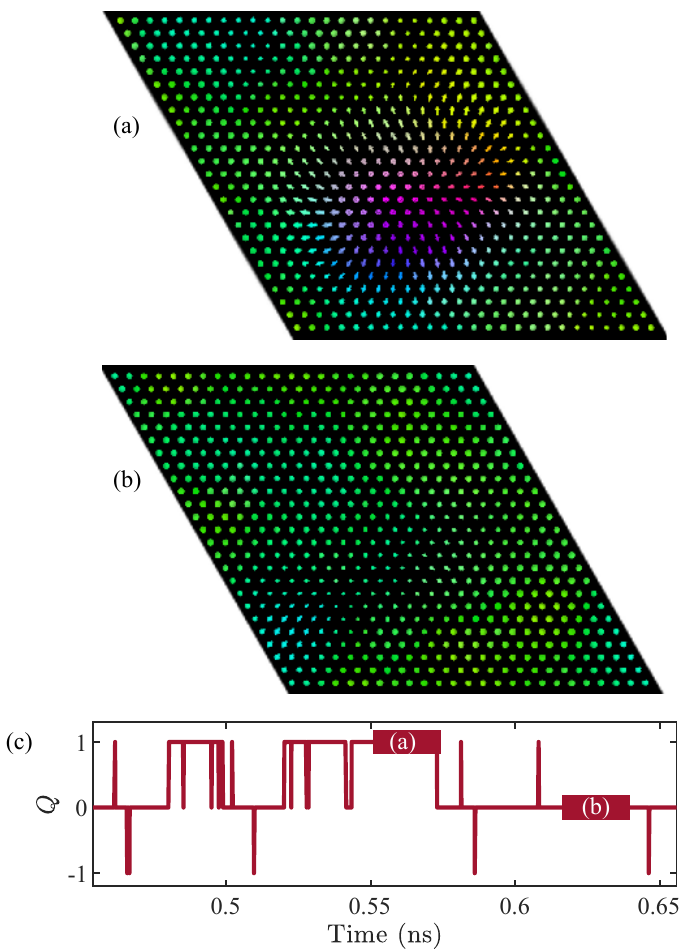}
\caption{Time-averaged spin configurations for topological charges (a) $Q=1$ and (b) $Q=0$. (c) Time evolution of the topological charge for a sample run with external field strength $B=1$~T, $T=80$~K and number of spins $N_{S} = 25 \times 25$. The spin configurations in (a) and (b) result from a time average for each spin's Cartesian coordinates over the indicated time intervals.}
\label{QoverT}
\end{figure}

As a first step, it has to established whether configurations with topological charge $Q=1$ may indeed be identified as skyrmions in our simulations, while $Q=0$ states remain close to the collinear configuration. This question is especially relevant in the strongly fluctuating regime above $T_{\textrm{c}}\approx50$~K, illustrated at $T=80$~K in Figs.~\ref{QoverT}(a) and (b). 
Besides the localized, cylindrically symmetric equilibrium skyrmion known at zero temperature, a $Q=1$ spin configuration may denote a combination of two skyrmions plus an antiskyrmion or a completely disordered state with various signs of the local topological charge density. However, the latter configurations turn out to be significantly higher in energy and are consequently expected to occur very rarely in our simulations for a wide temperature range. In Fig.~\ref{QoverT}(c), the topological charge is shown over a short timescale for a sample run with visible changes in the topological charge over time. We calculate a time average of each spin's Cartesian coordinates over the time intervals denoted by thick lines in Fig.~\ref{QoverT}(c)  in order to demonstrate that the recorded spin structures with $Q=1$ are actually skyrmions. Note that the individual spins do not have unit length after taking the average. Figure~\ref{QoverT}(a) demonstrates that the average structure still consists of a downwards-pointing core in an upwards-pointing background, with the spin directions spanning the whole unit sphere as indicated by the color-coding. The time average over configurations with $Q=0$, shown in Fig.~\ref{QoverT}(b), still resembles the collinear state even at this elevated temperature.

By determining the skyrmion lifetime as a function of temperature between $T=60$~K and $T=85$~K, an energy barrier of $\Delta E/k_{\textrm{B}}\approx 922$~K was obtained using the system parameters in Fig.~\ref{QoverT}.

\begin{figure}
\includegraphics[width=0.9\columnwidth]{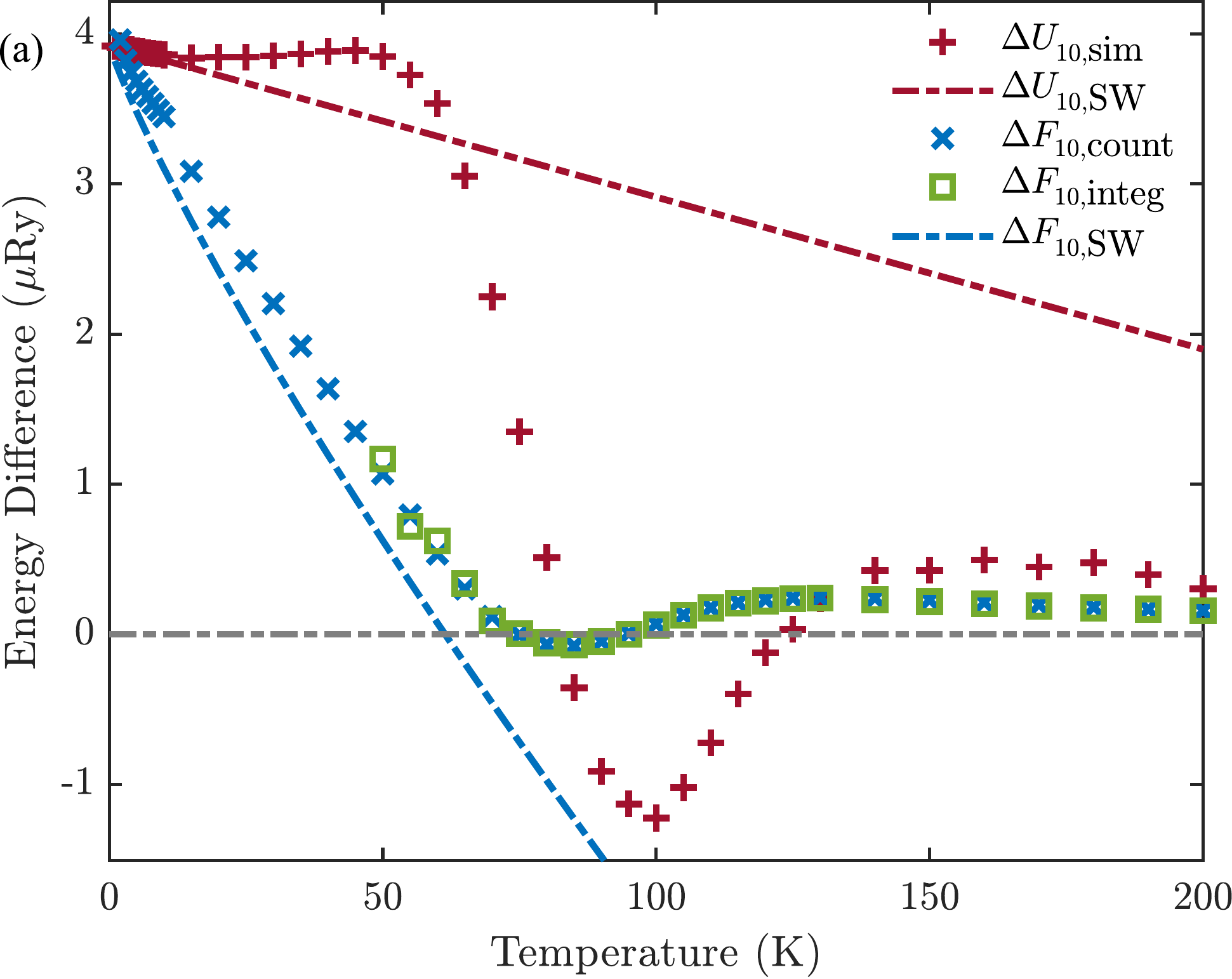}
\includegraphics[width=0.9\columnwidth]{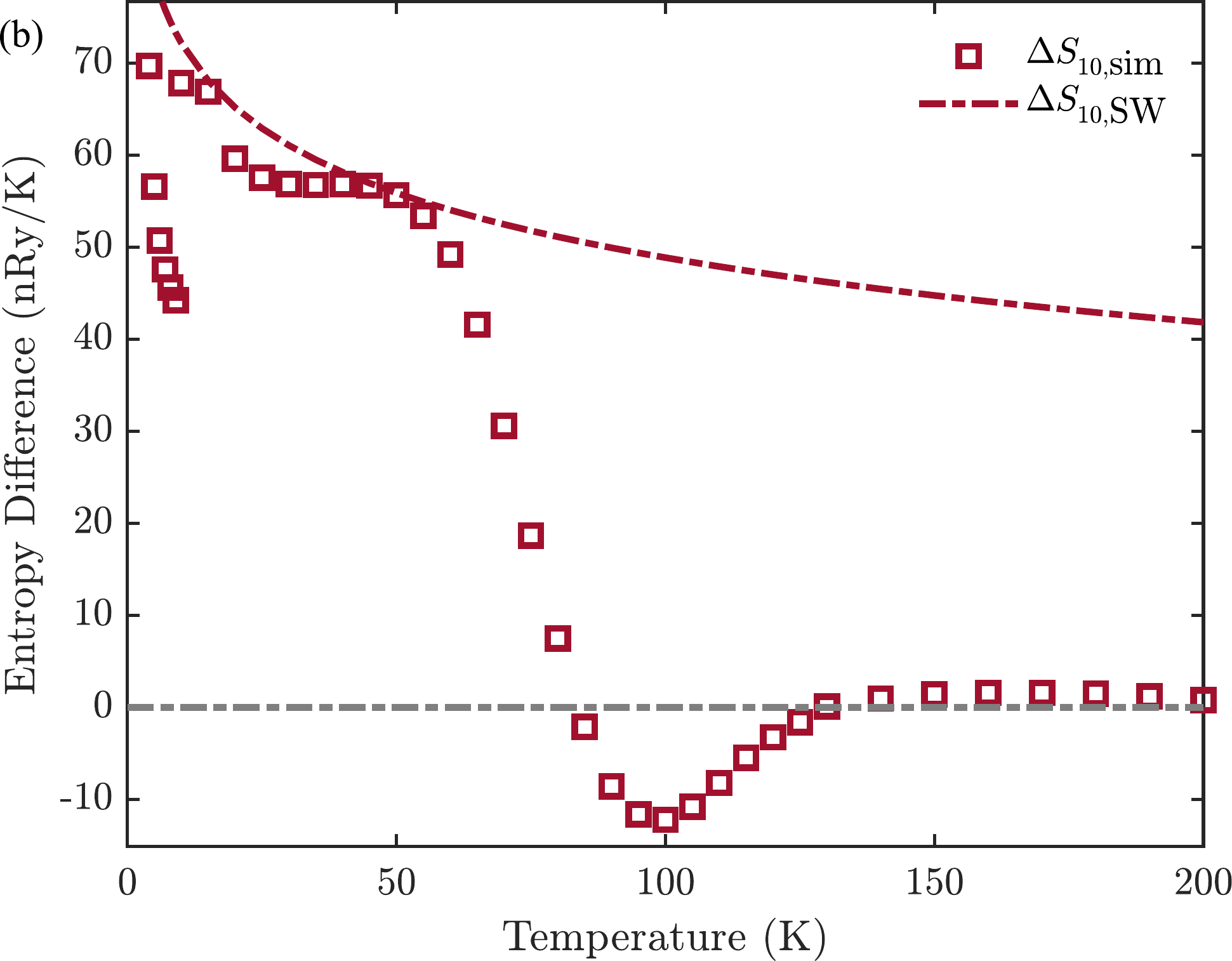}
\caption{(a) Internal- and free-energy difference per spin between skyrmion ($Q=1$) and topologically trivial ($Q=0$) states as a function of temperature, for $B=1$~T and $N_{S} = 25 \times 25$. The free-energy difference $\Delta F_{10,\textrm{count}}$ is calculated from Eq.~\eqref{DeltaFN} and $\Delta F_{10,\textrm{integ}}$ from Eq.~\eqref{DeltaFInteg}. (b) Difference in entropy per spin between skyrmionic and topologically trivial states as a function of temperature, for the same parameters. Dashed lines denote the prediction of linear spin-wave theory, Eq.~\eqref{eqn10}, \eqref{eqn11}, and \eqref{eqn13}.
}
\label{DeltaEF}
\end{figure}

The difference in internal and free energy between the $Q=0$ and $Q=1$ states can be seen in Fig.~\ref{DeltaEF}(a). For using Eq.~\eqref{DeltaFInteg}, $\Delta F_{10,\textrm{count}}$ was determined at a temperature of $T_0 = 190$~K by comparing the number of states with different topological charges. Because the free-energy differences agree between the counting and the integral methods for temperatures $T>70$~K , the choice of $T_0$ is not critical for our results. The deviations at lower temperature can be attributed to the limitations of the simulation length discussed after Eq.~\eqref{DeltaFN}.

Since the metastability field $B_{\textrm{m}}$ for skyrmions in Fig.~\ref{phasediag} is significantly lower than the value of $B=1$~T used in Fig.~\ref{DeltaEF},  at low temperatures the topologically trivial configuration 
is strongly preferred. $\Delta F_{10}$ has a minimum at around $T\approx85$~K with a value below $0$, showing that the skyrmion quasiparticles with the short lifetimes shown in Fig.~\ref{QoverT} are energetically preferred for a certain temperature range in this system even for such a high value of the external field. From Eq.~\eqref{DeltaFN}, it is clear that skyrmions occur more often at these temperatures than topologically trivial states. For higher temperatures, $\Delta F_{10}$ is slightly positive, but rapid changes in $Q$ may be observed in this regime. 

Up to $T\approx50$~K, the 
weak temperature dependence of the internal-energy difference and the linear decrease of the free-energy difference agree with the predictions of linear spin-wave theory in Sec.~\ref{sec2c}. The negative slope of $\Delta F_{10}$ indicates the entropic stabilization of skyrmions~\cite{Hagemeister2015, Desplat2018, Malottki2019}. 
Note that for the considered external field and system size, the free-energy difference only reaches negative values at higher temperature where deviations from linear spin-wave theory are observed, particularly in the rapid reduction of the internal-energy difference.

From $\Delta F_{10}$, we calculate $\Delta S_{10}$ using Eq.~\eqref{eqn9}, with the result shown in Fig.~\ref{DeltaEF}(b). Between $25$~K and $50$~K, the entropy difference is slightly decreasing, in agreement with Eq.~\eqref{eqn13} derived from linear spin-wave theory and indicated by the dashed line in the figure. Unfortunately, at very low temperature where the logarithmic dependence would dominate, the inaccuracies caused by the numerical integration in Eq.~\eqref{DeltaFInteg} and differentiation in Eq.~\eqref{eqn9} obscure its influence.

Above $50$~K, approximately corresponding to the critical temperature determined at zero field in Fig.~\ref{phasediag}(b), the entropy difference between the considered states is drastically reduced. As shown in the figure, spin-wave theory loses its validity in this regime due to the strong thermal fluctuations, which lead to the rapid creation and destruction of metastable skyrmions. Remarkably, the skyrmion quasiparticles actually have a lower entropy for certain temperatures than $Q=0$ states. The temperature range where $\Delta S_{10}$ is negative is confined between the extrema of $\Delta F_{10}$, as expected from the derivative expression~\eqref{eqn9}, and shifted towards higher temperatures as compared to the temperature range with negative $\Delta F_{10}$. It is established that in the low-temperature limit where spin-wave theory is applicable, the competition between positive internal-energy difference and negative entropy difference contributions causes the free energy of skyrmions to become lower than that of the collinear state as the temperature is increased~\cite{Muelbauer2009}. However, these results indicate that for skyrmion quasiparticles with reduced lifetimes in the strongly fluctuating regime 
the stabilization mechanism is more complex, and the role of the internal-energy and the entropy terms may become reversed as both of them change sign. In the skyrmion lifetime, the pre-exponential factor of the Arrhenius law is similarly affected by the entropy of the different states~\cite{Hagemeister2015, Desplat2018, Malottki2019, Bessarab2018,Wild2017},
meaning that the observed decrease in $\Delta S_{10}$ may also influence the lifetime in this regime. The temperature in this regime is still relatively low compared to the energy barrier $\Delta E/k_{\textrm{B}}\approx 922$~K, reinforcing the validity of the Arrhenius expression. 

\begin{figure}
\includegraphics[width=0.9\columnwidth]{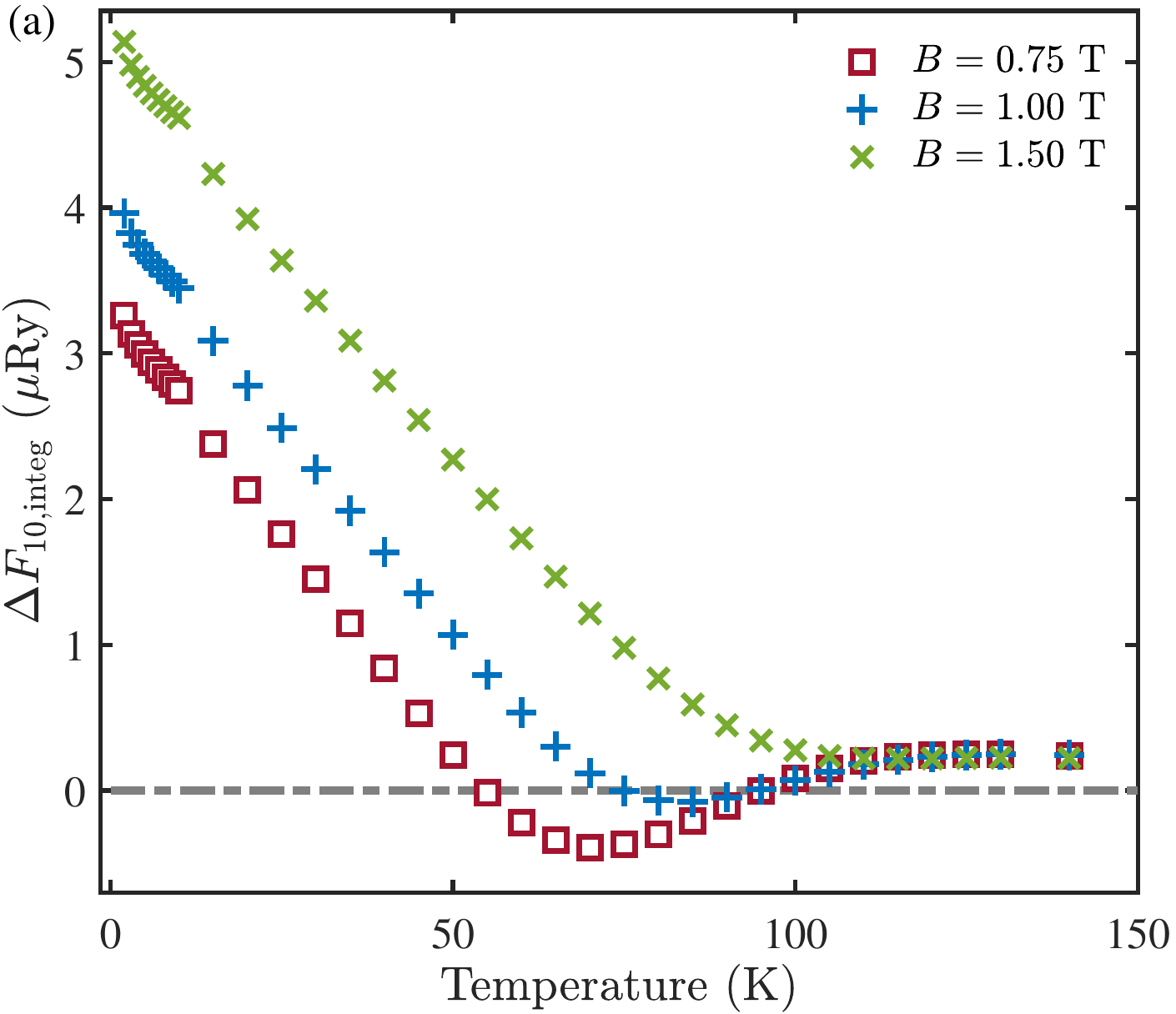} 
\includegraphics[width=0.9\columnwidth]{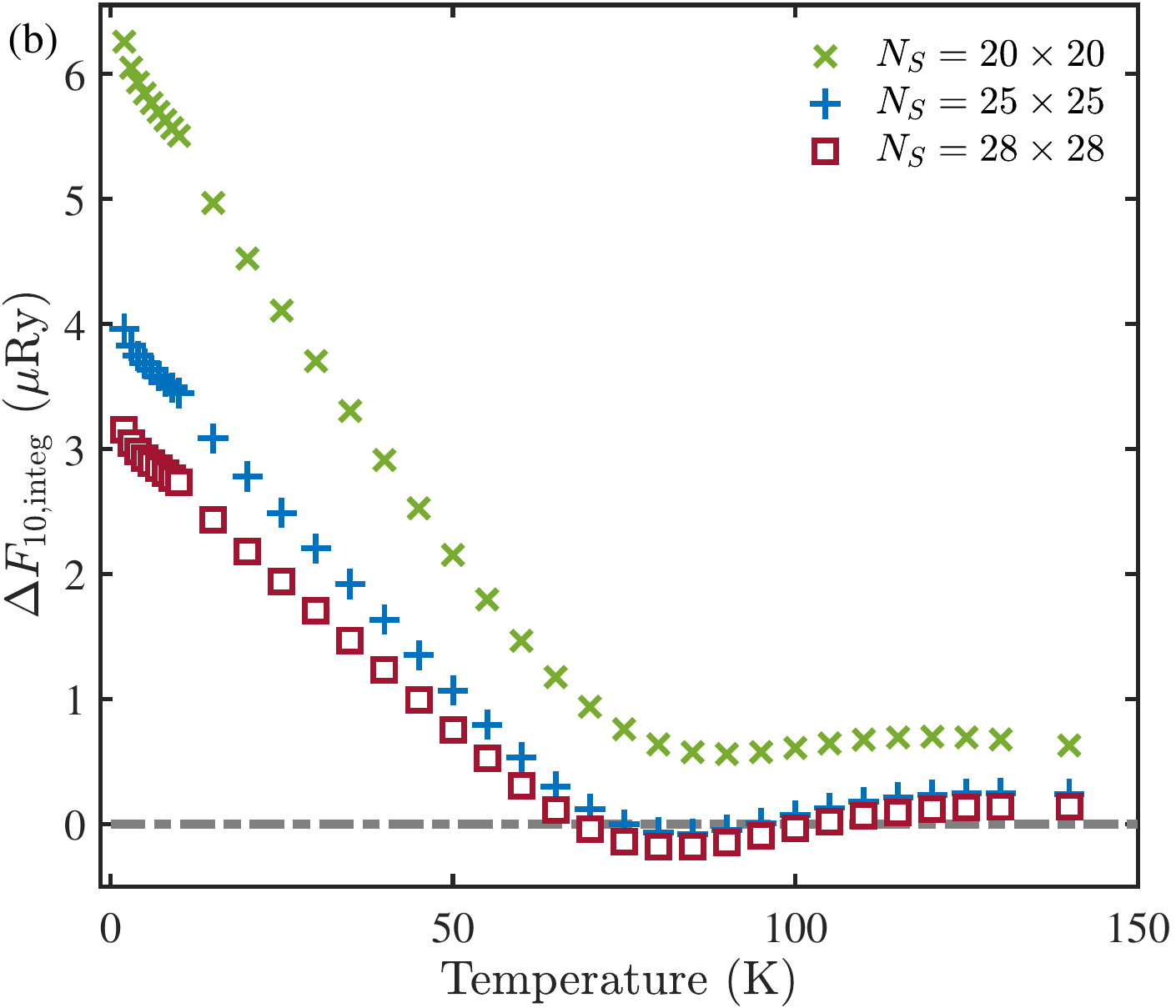}
\caption{(a) Free-energy difference per spin as a function of temperature for different magnetic fields at a simulated system size of $N_{S} = 25 \times 25$ spins. (b) Free-energy difference per spin as a function of temperature for different simulated system sizes at an external magnetic field of $B=1$~T. All curves are calculated using Eq.~\eqref{DeltaFInteg}.
}
\label{Compare}
\end{figure}

In Fig.~\ref{Compare}, $\Delta F_{10}$ is calculated for different external magnetic fields applied perpendicular to the surface and different sizes of the simulated system. It is visible in Fig.~\ref{Compare}(a) that strong external fields increase the internal- and free-energy difference at zero temperature, which can suppress the minimum in $\Delta F_{10}$, meaning that skyrmion quasiparticles cannot become favored even at elevated temperatures. At lower magnetic fields, the temperature range where skyrmions are energetically preferred is larger and it extends to lower temperatures. Also the minimum value of $\Delta F_{10}$ is even lower than for higher field values. The lower limit of the temperature range where skyrmions are stable is expected to reach $0$~K at $B_{\textrm{m}}\approx0.1$~T, where isolated skyrmions on an infinite collinear background become energetically preferable (cf. Fig.~\ref{phasediag}(a)).

Simulations with different system sizes are compared in Fig.~\ref{Compare}(b). In our finite-size system, the skyrmions interact with themselves via the periodic boundary conditions, thereby raising the internal-energy difference and with that also the free-energy difference. Since even isolated skyrmions have a higher internal energy at zero temperature than the collinear state, the free-energy per spin also decreases with increasing the system size for a fixed number of skyrmions as the relative size of the field-polarized areas increase. However, the increased ratio of the field-polarized areas decreases the entropy difference as well, as indicated by the reduced slopes of the free-energy curves in the low-temperature regime in Fig.~\ref{Compare}(b). It is obvious that simulating with too small systems can cause the minimum in $\Delta F_{10}$ to only have positive values, which means that topologically trivial states are always preferred over $Q=1$ states. On the other hand, increasing the number of simulated spins lowers the free-energy difference per spin, as can be seen by comparing the case $N_{S}=25 \times 25$ to $N_{S}=28 \times 28$, making $Q=1$ states also being preferred over a wider temperature range. Note that further increasing the system size may cause the formation of multiple skyrmions in the system at elevated temperature with a high probability, which effect was to be avoided in our simulations, similarly to Ref.~\cite{Hagemeister2015}. We mention that increasing the system size does not eliminate the interactions between skyrmions: at high temperature, a finite density of interacting skyrmions is observable instead of an isolated skyrmion at zero temperature, which is reflected as a finite probability of finding a skyrmion in the small systems considered here.

A numerical comparison of the entropy differences obtained from the simulations and from linear spin-wave theory at $T=25$~K is presented in Table~\ref{table1}. 
Treating the translational degrees of freedom as Goldstone modes following Eq.~\eqref{eqn13} provides reasonable agreement with the simulation data. This is expected since the energy barrier between different positions of the skyrmion created by the modulating potential of the atomic lattice, which technically breaks the continuous translational symmetry, is negligible at all simulated temperatures ($\Delta E/k_{\textrm{B}}\approx 10^{-8}$~K based on GNEB calculations). If the finite frequency of the translational modes is taken into account, similarly to the procedure in Refs.~\cite{Desplat2018,Desplat2020}, then achieving agreement with the simulation results requires assuming a frequency which is around three orders of magnitude higher than the numerically calculated value for the translation mode. Treating the translations as a finite-frequency mode does not reproduce the weak logarithmic temperature dependence, which is more pronounced for smaller system sizes, in agreement with 
Eq.~\eqref{eqn13}; see Supplemental Fig.~1~\cite{supp}.

Linear spin-wave theory apparently reproduces the decrease of the entropy difference for larger system sizes observable in Fig.~\ref{Compare}(b) and discussed above. While the entropy difference only depends weakly on the external field in the considered regime, it is remarkable that the linear regime in the free-energy difference, indicative of the validity of the linear spin-wave approximation, extends over a larger temperature range for higher field values. On the one hand, this effect is rather counterintuitive when only considering the energy barriers, namely the energy differences between the stable states and the saddle point configuration $S$: $\Delta E_{S1}$ for skyrmion annihilation decreases as the field is increased, while $\Delta E_{S0}$ for skyrmion creation stays mostly constant (see Table~\ref{table2} for the numerical values), which would point towards a reduced skyrmion stability and less reliability of the linear spin-wave approximation. On the other hand, the spin-wave frequencies also increase for higher field values, as shown in Ref.~\cite{Rozsa_2020} for the present system. This means that although the energy barriers become lower as the field is increased, the minima simultaneously become sharper, and less magnons are excited at the same temperature. The reduced number of magnons suppresses the interaction between them, which may explain why the non-interacting linear spin-wave model remains applicable in a wider temperature range. This observation also agrees with the phase diagrams obtained in Refs.~\cite{Rozsa2016,Boettcher2018}: the crossover temperature from the field-polarized regime to the completely disordered paramagnetic regime is increasing for higher external field values, and this temperature value marks the maximum of spin fluctuations at a fixed field value.

\begin{table}
 \caption{\label{table1} Entropy difference between the $Q=1$ and $Q=0$ states at $T=25$~K, obtained from numerical simulations 
and from linear spin-wave theory Eq.~\eqref{eqn13}, respectively.
 }
\begin{ruledtabular}
 \begin{tabular}{cccc}
   $N_{S}$                        & $B$~(T)                       & $\Delta S_{10,\textrm{sim}}$~(nRy/K) & $\Delta S_{10,\textrm{SW}}$~(nRy/K) \\\hline
   $20\times20$     & 1.00 & 82.05 & 91.78  \\ 
   $25\times25$     & 0.75 & 61.08 & 64.38  \\ 
   $25\times25$     & 1.00 & 57.55 & 62.92  \\ 
   $25\times25$     & 1.50 & 56.05 & 63.09  \\ 
   $28\times28$     & 1.00 & 47.42 & 52.08  \\ 
 \end{tabular}
\end{ruledtabular}
\end{table}

\begin{table}
 \caption{\label{table2} Energy barriers for skyrmion annihilation $\Delta E_{S1}$ and skyrmion creation $\Delta E_{S0}$, where $S$ denotes the saddle-point configuration between the $Q=1$ and $Q=0$ states. The annihilation barrier was determined from an Arrhenius fit to the skyrmion lifetimes obtained from the simulations in the range $60~\textrm{K}\le T\le 85~\textrm{K}$, while for the creation barrier the relation $\Delta E_{S0}=\Delta E_{S1}+\Delta E_{10}$ was used. Values obtained from the GNEB method are provided for comparison, displaying the same trends when varying the system size and field strength.
 }
\begin{ruledtabular}
 \begin{tabular}{cccccc}
   $N_{S}$                       & $B$~(T)                       & \multicolumn{2}{c}{$\Delta E_{S1}/k_{\textrm{B}}$~(K)} & \multicolumn{2}{c}{$\Delta E_{S0}/k_{\textrm{B}}$~(K)} \\
    & & sim & GNEB & sim & GNEB \\\hline
   $20\times20$     & 1.00 & 918  & 800 & 1306 & 1190 \\ 
   $25\times25$     & 0.75 &  1008 & 850 & 1328 & 1170 \\ 
   $25\times25$     & 1.00 &  922 & 800 & 1310 & 1190 \\
   $25\times25$     & 1.50 & 835  & 720 & 1335 & 1220 \\ 
   $28\times28$     & 1.00 & 929  & 800 & 1321 & 1190 \\ 
 \end{tabular}
\end{ruledtabular}
\end{table}

\begin{figure}
\includegraphics[width=0.9\columnwidth]{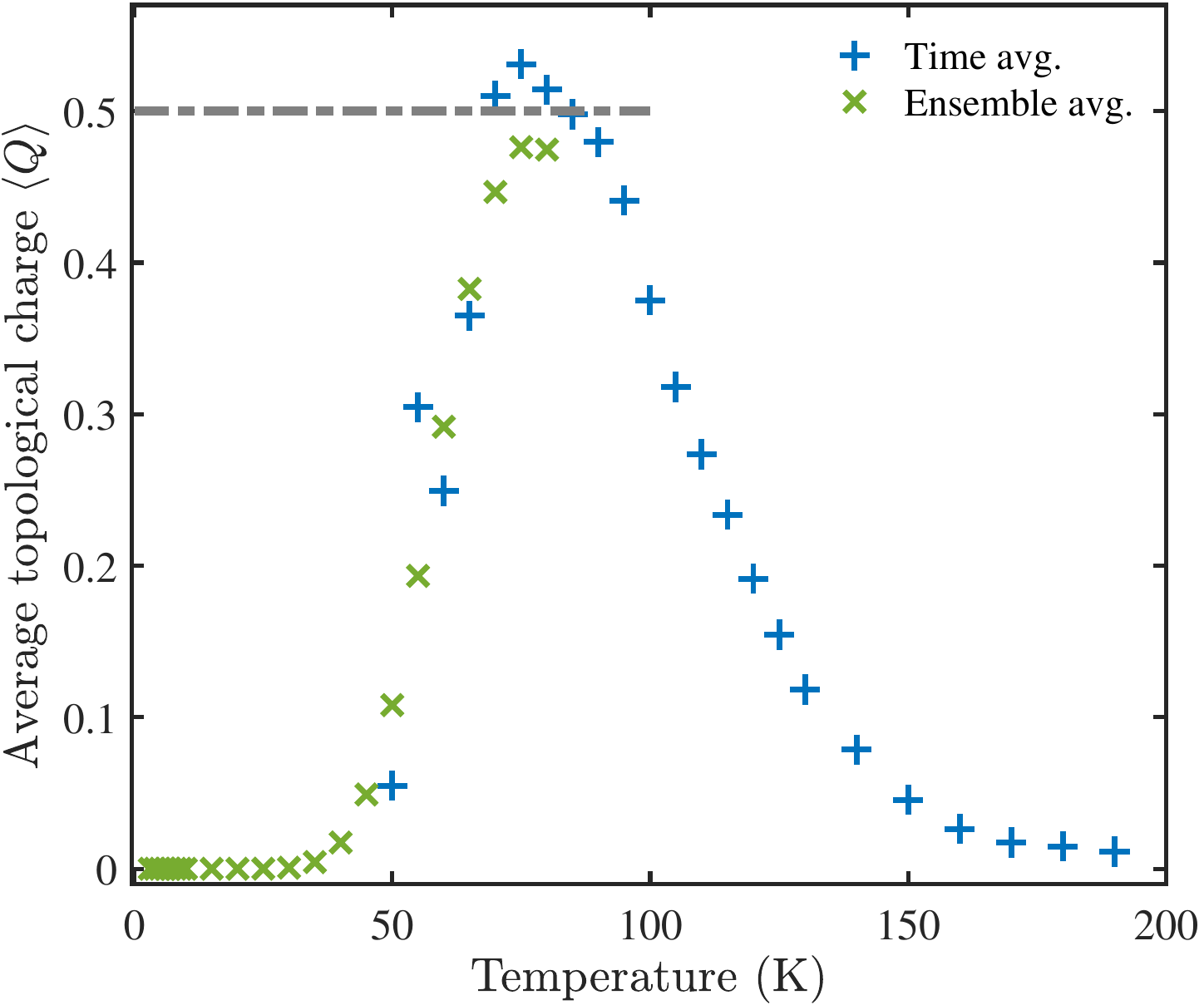}
\caption{Average value of Q calculated with ensemble average using free-energy values of topological charges $Q=-1, 0, 1$ (green) and time average over all topological charges (blue). System simulated with $N_{S}=25 \times 25$ spins and $B = 1$~T external field.
}
\label{pQ}
\end{figure}

At high temperatures, the topological charge can take many different values during a single simulation, and the time average in Eq.~\eqref{Qtime} was used for calculating the average topological charge. The resulting average of the topological charge $\left<Q\right>$ as a function of temperature $T$ is shown in Figure~\ref{pQ}. At low temperature where the transition times between different topological charges exceeds the simulation times, the ensemble average in Eq.~\eqref{Qens} was approximated by using the free energies of topological charges $Q=-1, 0$, and $1$. For the data with topological charge $Q = -1$ we performed simulations with an antiskyrmion as the initial condition. Since the antiskyrmion has considerably higher energy than the skyrmion in the system~\cite{Rozsa2017}, we found no minimum either in the free-energy difference $\Delta F_{-10}$ or in the entropy difference $\Delta S_{-10}$; see Supplemental Fig.~2~\cite{supp}.

In Fig.~\ref{pQ} it can be observed that the thermal fluctuations cause a considerable increase in the average topological number in the system, especially in the regime directly above $T_{\textrm{c}}\approx50$~K. This is in agreement with the results of Refs.~\cite{Rozsa2016,Boettcher2018} in skyrmionic systems similarly described by spin models based on \textit{ab initio} calculations. Notice that at around $T\approx85$~K, there is a regime with average topological charge above $0.5$ for the time average. This provides further evidence that skyrmions are thermodynamically preferred at elevated temperatures in the considered system. Note that the ensemble average is close to the time average between $50$ and $80$~K, but it stays below $0.5$ for all temperature values. Although the $Q=1$ state has lower free energy than the collinear state close to the maximum of the average topological charge, taking the presence of antiskyrmions into account reduces the average value below $0.5$. However, a lot of changes of the topological charge are recorded at this temperature range, as can be seen in Fig.~\ref{QoverT}, and considering higher $Q$ values in the time average increases the average above $0.5$. At low temperature, the average topological charge vanishes since skyrmions are energetically unfavorable compared to the topologically trivial state. At very high temperature, $\left<Q\right>$ again converges to zero since all microstates, including those with positive and negative topological charges, contribute with similar weights to the total partition function in this limit. Figure~\ref{pQ} demonstrates that combining Eq.~\eqref{Qtime} accurate at high temperature with Eq.~\eqref{Qens} which can be applied at low temperature enables the calculation of the average topological charge ranging from the completely ordered to the completely disordered regime, with reasonable agreement between the two methods in the intermediate temperature range where both of them are valid.

\section{Conclusion}
We calculated the free-energy and entropy difference between a skyrmion and the topologically trivial state in a (Pt$_{0.95}$Ir$_{0.05}$)/Fe bilayer on a Pd(111) surface by means of numerical simulations. We found that the free-energy difference turns from positive to negative as the temperature is increased, meaning structures with $Q=1$ are thermodynamically preferred over topologically trivial states in a certain temperature range. We demonstrated that this range vanishes at higher magnetic fields or smaller system sizes, where the internal energy of skyrmions with respect to the collinear state becomes higher. We showed that $Q=1$ configurations in a time average can still be identified with skyrmion-like spin structures in this temperature range, although they are frequently created and destroyed by thermal fluctuations. The preference for the formation of skyrmions at elevated temperature agrees with the prediction of entropic stabilization based on linear spin-wave theory, but qualitative deviations from this approximation have been observed in the thermodynamic quantities. In particular, we found that while skyrmions have higher entropy at low temperature, their presence reduces the entropy at elevated temperatures. We calculated a composite average of the topological charge via combining an approximate average in the canonical ensemble based on the free-energy calculations at low temperature with a time average at higher temperatures. We found the time average of the topological charge to reach values over $0.5$ in the temperature range where we found skyrmions to be thermodynamically preferred.

Although skyrmions are preferred by the free energy in a certain parameter regime, this does not mean that these topologically non-trivial states are stable at this temperature. The deviations from linear spin-wave theory based on stable equilibrium structures are pronounced in this regime characterized by strong thermal fluctuations, and the lifetime of skyrmions is considerably reduced as confirmed by our simulations. This shows that skyrmions cannot be interpreted as particles with a conserved topological charge, but should rather be seen as quasiparticles with a finite chemical potential. The non-integer average topological number in this regime corresponds to the probability of finding a skyrmion in the system, if the contributions from higher or opposite topological charges can be neglected. These results should stimulate further studies on the properties of topologically non-trivial spin structures in the presence of strong thermal fluctuations.

\begin{acknowledgments}

The authors would like to thank Bertrand Dup\'{e} for stimulating discussions. Financial support by the German Research Foundation via SFB 1432 and via Project No. 403502522 and by the National Research, Development and Innovation Office of Hungary via Project No. K131938 is gratefully acknowledged.

\end{acknowledgments}

\end{document}